# Borophene as an anode material for Ca, Mg, Na or Li ion storage: A first-principle study

Bohayra Mortazavi[*,1], Arezoo Dianat[2], Obaidur Rahaman[1], Gianaurelio Cuniberti[2,3,4], Timon Rabczuk[1,#]

[1]Institute of Structural Mechanics, Bauhaus-Universität Weimar, Marienstr. 15, D-99423 Weimar, Germany.

[2]Institute for Materials Science and Max Bergman Center of Biomaterials, TU Dresden, 01062 Dresden, Germany

[3]Dresden Center for Computational Materials Science, TU Dresden, D-01062 Dresden, Germany

[4]Center for Advancing Electronics Dresden, TU Dresden, 01062 Dresden, Germany

## Abstract

Borophene, the boron atom analogue to graphene, being atomic thick have been just recently experimentally fabricated. In this work, we employ first-principles density functional theory calculations to investigate the interaction of Ca, Mg, Na or Li atoms with single-layer and free-standing borophene. We first identified the most stable binding sites and their corresponding binding energies as well and then we gradually increased the ions concentration. Our calculations predict strong binding energies of around 4.03 eV, 2.09 eV, 2.92 eV and 3.28 eV between the borophene substrate and Ca, Mg, Na or Li ions, respectively. We found that the binding energy generally decreases by increasing the ions content. Using the Bader charge analysis, we evaluate the charge transfer between the adatoms and the borophene sheet. Our investigation proposes the borophene as a 2D material with a remarkably high capacity of around 800 mAh/g, 1960 mAh/g, 1380 mAh/g and 1720 mAh/g for Ca, Mg, Na or Li ions storage, respectively. This study can be useful for the possible application of borophene for the rechargeable ion batteries.

*Corresponding author (Bohayra Mortazavi): bohayra.mortazavi@gmail.com
Tel: +49 157 8037 8770, Fax: +49 364 358 4511,#Timon.rabczuk@uni-weimar.de

## 1. Introduction

Two-dimensional (2D) materials are currently among the most interesting research topics because of their remarkably wide application prospects from nanoelectronics to aerospace structures. This interest was principally initiated by the successful production of graphene [1–3] which is the planar arrangement of carbon atoms with honeycomb lattice. Graphene is a semi-metallic material that exhibits outstanding mechanical [4] and heat conduction [5] properties superior than all known materials. Graphene success motivated numerous experimental and theoretical researches toward the prediction and synthesis of other two-dimensional (2D) compounds, such as hexagonal boron-nitride [6,7], silicene [8,9], germanene [10], stanene [11] and transition metal dichalcogenides [12–14] like molybdenum disulfide ($MoS_2$).

In line with the continuous advances in the fabrication of 2D materials, an exciting development has just taken place with respect to the synthesis of 2D boron films, so called borophene [15]. Interestingly, the existence of borophene and its metallic properties were already theoretically predicted [16,17]. In response to the high demand for more efficient rechargeable energy storage systems, graphene and other 2D crystals and their hybrid structures can be investigated as promising candidates owing to their large surface area, high mechanical flexibility and high electron mobility [18–21]. For example, bulk silicon anode can yield a specific capacity of 4200 mAh/g [22,23], but due to the degradation by drastic volume changes during the battery operation the commercialization of this battery technology is prohibited so far [24–26]. On the other side, silicene the planar form of silicon can exhibit high capacity without being destroyed during lithiation [21]. In addition, $MoS_2$ nanosheets can yield a high capacity of 912 mAh/g at 1C current [27]. One of the most attractive characteristics of the 2D materials is related to their potential for integration and fabrication of heterostructures [28,29] with tuneable properties. Because of experimental complexities at nanoscale, theoretical investigations can be used as an alternative to evaluate the application prospect of the 2D materials [30–38]. This study therefore aims to explore the applicability of borophene for Ca, Mg, Na or Li ions storage using first-principles density functional theory calculations.

## 2. Modelling

First-principles density functional theory (DFT) calculations in the present study were carried out using Vienna ab initio simulation package (VASP) [39,40] using the Perdew-Burke-Ernzerhof (PBE) generalized gradient approximation exchange-correlation functional [41]. We considered van der Waals interactions using the semi-empirical correction of Grimme [42,43]. The projector augmented wave method [44] was employed with an energy cutoff of 500 eV. We constructed a borophene sheet including 72 atoms and applied periodic boundary conditions in all directions with a 20 Å vacuum layer to avoid image-image interaction along the sheet thickness. We first identified the most favorable binding sites for Ca, Mg, Na or Li adatom on the borophene film. Then we gradually and step by step increased the atoms concentration by randomly but uniformly positioning them on the identified sites. After automatically positioning the ions, conjugate gradient method energy minimization was performed on the whole structure at each step with a $10^{-4}$ eV criteria for energy convergence using a 7×7×1 k-point mesh size for Brillouin zone sampling. A single point calculation was then performed at the end of each minimization step to report the energy of the system in which the Brillouin zone is sampled using a 8×8×1 k-point mesh size. For the evaluation of charge transfer between the ions and the substrate, we performed Bader charge analysis [45]. Climbing-image nudged elastic band (NEB) [46] method was utilized to obtain the diffusion pathways and corresponding energy barriers. For the NEB calculations, we included 24 atoms for the borophene film and a Monkhorst Pack mesh [47] of 15×15×1 special points was used for integration in the reciprocal space. Ab initio molecular dynamics (AIMD) simulations were also performed using Langevin thermostat with a time step of 1fs and 2×2×1 k-point mesh.

## 3. Results and discussions

We first investigate the adsorption behaviour of a single Ca, Mg, Na or Li atoms on the borophene sheet. For a material to be suitable as an anode for a particular ion storage, a relatively large energy for the ion adsorption plays a fundamental role. To find the site with the strongest binding energy for a particular ion we considered four different initial adsorption sites: above the top of boron atoms in the furrow, above the middle-point of B–B bond in the furrow, above the top of boron atoms in the ridge and finally above the middle-point of B–B bond in the ridge [48]. In this case, next we performed energy minimization using the conjugate gradient method with a strict energy convergence criteria of $10^{-6}$ eV. In Fig. 1, top and side views of a single Li or Na ions adsorbed on the borophene sheet are illustrated. We found that for Li atom the most stable site for the adsorption is above the top of boron atoms in the furrow (Fig. 1a), which is in agreement with the recent theoretical study [48]. Nevertheless, for Ca, Mg and Na atoms we found that the most stable position for the ion adsorption is above the middle-point of B–B bond in the furrow (Fig. 1b).

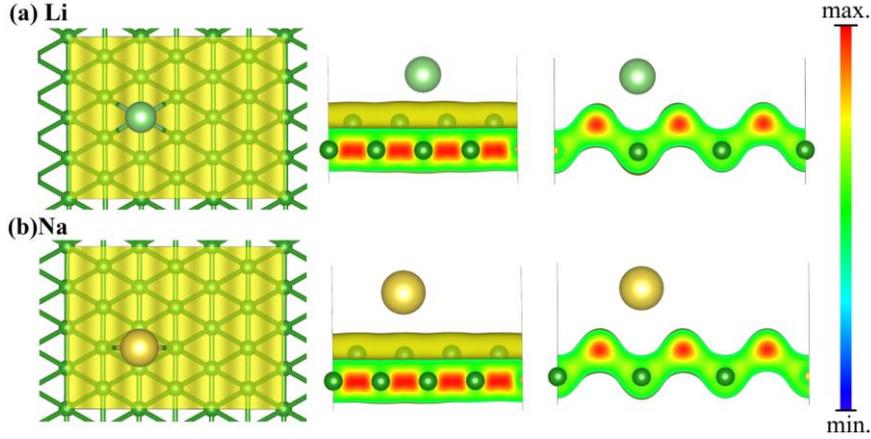

Fig. 1, Top and side views of the most stable site for the adsorption of a single (a) Li or (b) Na atoms on the borophene. For the borophene sheet we plotted the valence electron density. For Mg and Ca atoms the most stable site for the adsorption is similar for Na ion which is the middle-point of B–B bonds in the furrow. VESTA [49] package was used for the illustration of structures.

During the charging and discharging processes of a rechargeable battery, the ions concentration increases or decreases depending on the current direction. These processes consequently induce compositional and structural change of the anode material. In the next step we therefore investigate the effects of gradual increase of the atom content on the borophene anode. In this case, in order to be close to the minimized structure, we randomly and uniformly positioned the ions on the stable sites that were predicted in the previous step. After performing the energy minimization, we again increased the ions concentration. In the present study, to have a good statistical view, we performed four independent simulations for the ions adsorption on the borophene. After the every step of atoms insertion and geometry optimisation, we calculated the adsorption energies and performed the Bader charges analysis. The charge capacity of borophene for a particular ion is correlated to the state at which any further increasing of the atom concentration does not transfer additional electrons to the substrate. Based on the Bader charges analysis, we obtained the maximum charge capacity for the considered atoms on the borophene anode. In Fig. 2, we illustrate samples of borophene films saturated with Ca, Mg, Na or Li ions. Our calculations suggest remarkably high capacity of around 800 mAh/g, 1960 mAh/g, 1380 mAh/g and 1720 mAh/g for Ca, Mg, Na or Li ions adsorption on borophene, respectively. Our prediction for the capacity of Li ion storage on borophene is around 7% smaller than that predicted by a recent theoretical study [48]. Such a small discrepancy can be attributed to the difference in the modelling approach. In our study we included a relatively large borophene sheet in which ions adsorption occur randomly. Despite the statistical variations associated to such a modelling approach, that is more likely to what that happens during a real charging process of an anode material. Interestingly, our simulation results suggests that the Mg ions can yield the highest charge capacity among the considered atoms for a borophene anode material.

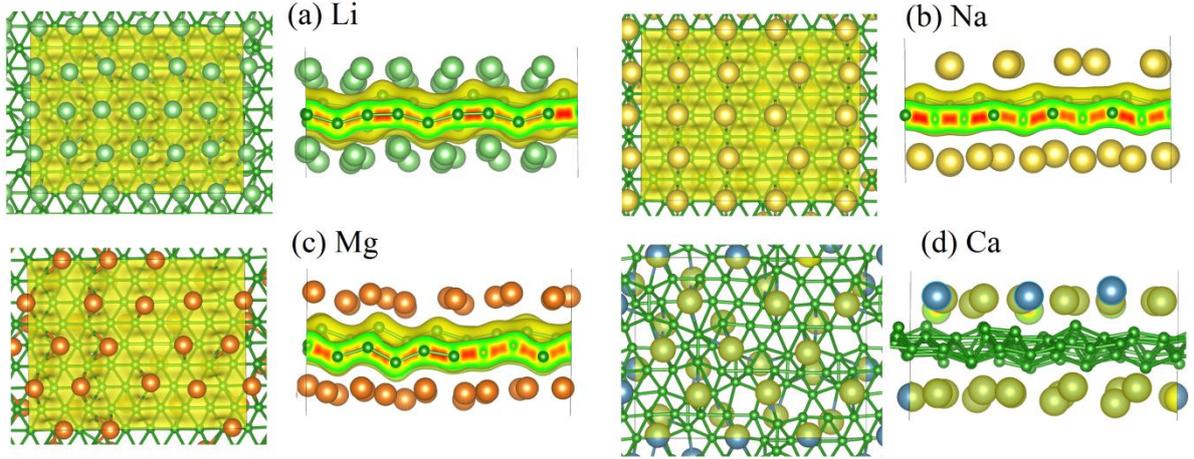

Fig. 2, Samples of borophene films saturated with (a) Li, (b) Na, (c) Mg and (d) Na ions. In our modelling Ca ions contain 10 ions in their valance bond and they only transfer a small portion of electrons to the borophene substrate. We predicted charge capacity of 800 mAh/g, 1960 mAh/g, 1380 mAh/g and 1720 mAh/g for Ca, Mg, Na or Li ions adsorption on the borophene, respectively.

Our modelling results illustrated in Fig. 3, reveals that high concentration of adatom adsorption induces considerable deformations and lattice distortions in the borophene substrate. We found the occurrences of the maximum and minimum distortions in the borophene lattice induced by Ca and Na atoms, respectively. For the application of borophene in rechargeable ion batteries these lattice deflections and distortions should be removable after the ions separation during the battery discharging process. To assess the reversibility of these structural deformations, we used AIMD simulations at room temperature. In this case, we removed the adatoms and then performed the AIMD simulations. We found that after 1 ps, 4 ps and 7 ps, the borophene sheets that were saturated by Na, Li and Mg ions, respectively, could well retrieve the original structures at 300 K. For the borophene sheets saturated with Ca atoms we found the formation of B–B bonds connecting the two neighbouring ridge and forming two pentagon rings. Based on the theoretical predictions these kind of bonds exist in Pmmm [16] borophene structure. To remove these bonds and retrieve the original structure, the borophene film that was covered by the Ca atoms needs a much longer relaxation time. Because of the capability of a considerably fast recovery of the original structure, the application of borophene for Na, Li or Mg ions storage is therefore expected to be practical. We also studied the thermal stabilities of the borophene structure at different temperatures using the AIMD method. We found that the borophene sheet remained intact at the end of the simulation at T=750 K. It was partly disintegrated at T=1000 K and T=1500 K and completely disintegrated at T=2000 K. The thermal stability of borophene at a relatively high temperature of 750 K is also a promising characteristic for an anode material.

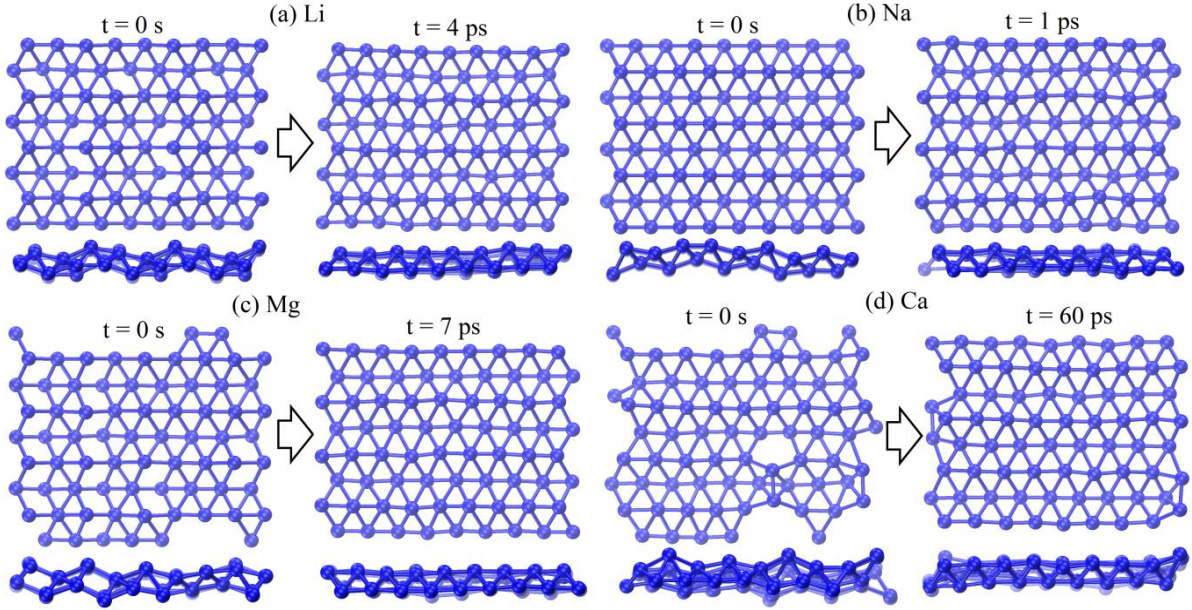

Fig. 3, AIMD simulation results to assess the irreversibility of borophene sheets deformations induced by the (a) Li, (b) Na, (c) Mg or (d) Ca, ions adsorption.

For a desirable anode material, presenting a high electronic conductivity is a property that directly affects the performance of the battery. Electronic conductivity is the main factor that determines the internal electronic resistance in a battery. In addition, the ohmic heating generated during the battery operation is also proportional to the electronic conductivity of the electrode's solid particles used for ions storage. Worthy to remember that predicted pristine borophene sheets exhibit various structural polymorphs that they all are metallic [15,17]. In this work, minimized borophene structures covered by Ca, Mg, Na or Li ions were therefore selected for the electronic density of states (DOS) calculation. Our calculated total DOS confirms that in all cases, despite of some structural deformations, the borophene films demonstrated metallic behaviour as indicated by the lack of any band gaps in the DOS.

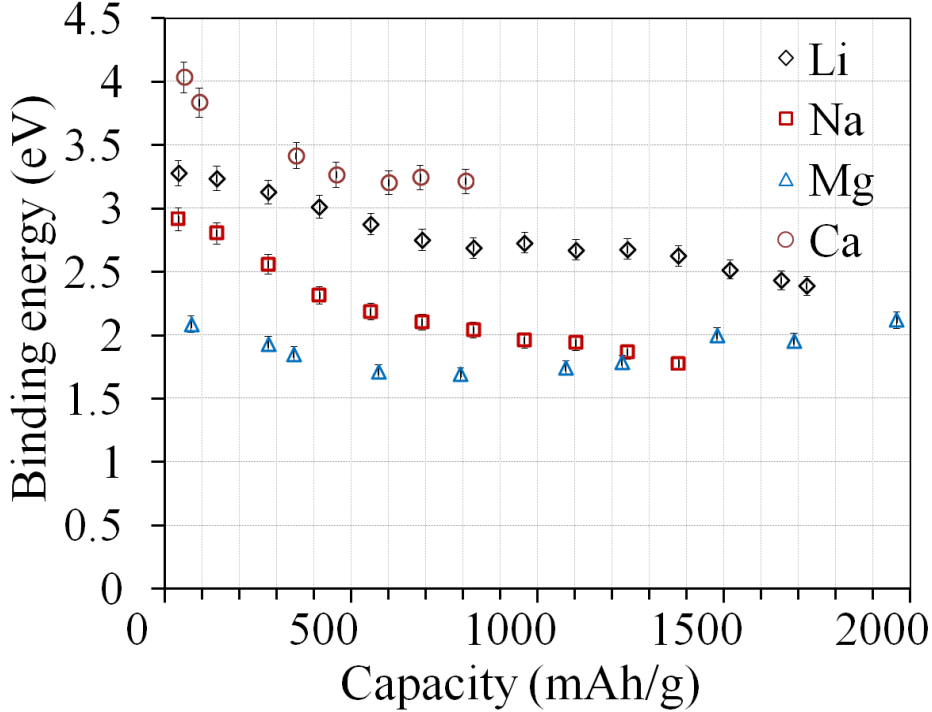

Fig. 4, Binding energy of Li, Na, Mg and Ca ions on borophene as the function of capacity.

We then discuss the binding (adsorption) energy of Ca, Mg, Na or Li atom on single-layer and free-standing borophene. The binding energy, $E_b$, is calculated using the following relation:

$$E_b = \frac{(E_{bor} + N \times E_{ion} - E_{bor+ion})}{N} \qquad (1)$$

where $E_{bor+ion}$ denotes the total energy of the structure, $E_{bor}$ is the total energy of the pristine borophene layer, $E_{ion}$ is the energy of an isolated atom and N is the number of adatoms. We predict a binding energy of 3.28 eV for a single Li ion adsorption on a borophene substrate. In addition, we predicted a relatively strong binding energies of around 4.03 eV, 2.09 eV and 2.92 eV between the borophene substrate and Ca, Mg and Na atoms, respectively. In Fig. 4, binding energies of Li, Na, Mg and Ca ions on borophene as the function of charge capacity are illustrated. In this case, as a general trend by increasing the Li, Na and Ca atoms concentration on borophene the binding energy decreases. On another hand, the binding energy of Mg atoms on the borophene initially slightly decreases and then almost remains constant by increasing the charge capacity.

Another factor that plays a significant role in the performance of electrode materials for rechargeable ion batteries is the rate of ions diffusion. Diffusion of ions inside the electrodes contributes directly to the charging or discharging loading rates accessible by a battery. In this regard, a faster ions diffusion is always a more desirable design parameter. We accordingly next shift our attention to the diffusion pathway of a single Ca, Mg, Na or Li ions on borophene. We used the NEB method to study the diffusion pathway as well as energy barrier of a single Ca, Mg, Na or Li adatom on single-layer borophene. This was achieved by calculating the variation in energy as ion moves between equivalent adsorption sites. The diffusion coordinate is the cumulative sum of the trajectory length of all atoms in the structure since the diffusion process involves cooperative motion of several atoms simultaneously, including the Ca, Mg, Na or Li adatom and its boron neighbours [21]. Because of the highly symmetric structure of borophene, two representative diffusion pathways were considered that are depicted in Fig. 5, one is the longitudinal path which is along the furrow (Fig. 5a) and the other is transverse path which is perpendicular to the furrow passing over the ridge (Fig. 5b). Based on our DFT results the diffusion energy barrier along the longitudinal and transverse pathways were predicted to be around 25 meV and 317 meV, respectively. Our prediction for the Li atom barrier energy along the transverse direction is close to the energy of 325.1 meV reported by Jiang *et al.* [48]. Nonetheless, our predicted value for longitudinal diffusion barrier energy is almost an order of magnitude higher than that obtained by Jiang *et al.* [48]. According to our simulation results, among the considered adatoms on the borophene anode, Na ions present the lowest energy barriers of 3 meV and 223 meV along the longitudinal and transverse directions, respectively. The Mg atom that was earlier predicted to yield the highest capacity, also have low diffusion energy barriers of 28 meV and 573 meV along the longitudinal and transverse directions, respectively. Regarding the diffusion barrier, again Ca adatoms on borophene present the weakest performance with longitudinal and transverse diffusion barriers of 440 meV and 612 meV, respectively. As expected because of the particular structure of the borophene sheet, for the all cases the energy barrier for diffusion along the longitudinal direction is conspicuously lower than that along the transverse direction. The considerably lower diffusion energy barrier along the longitudinal direction clearly confirms that the diffusion of Na, Mg and Li ions will occur totally along the furrow direction and the probability of diffusion along the transverse direction will be limited. As discussed in the work by Jiang *et al.* [48], and also in accordance with our results, the borophene films present very low ions diffusion energy barriers in comparison with other 2D materials [48].

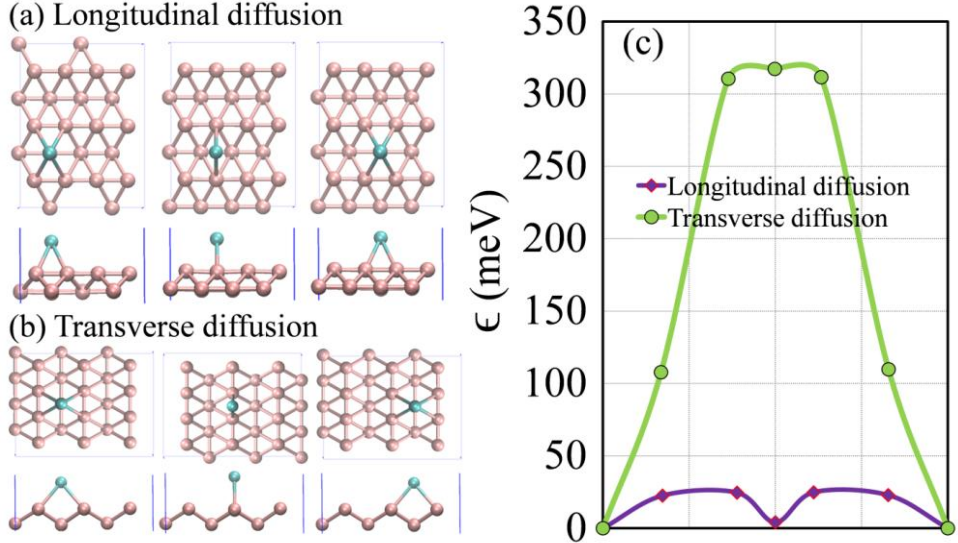

Fig. 5, Top and side views of a single Li atom diffusion pathways along (a) longitudinal and (b) transverse directions. (c) Calculated energy profile of lithium diffusion along longitudinal and transverse directions.

## 4. Conclusions

In summary, first-principles calculations were performed to probe the potential application of borophene for Ca, Mg, Na or Li ion storage. We first investigated strongest adsorption sites for a single adatom on the borophene film. We predicted relatively strong binding energies of around 4.03 eV, 2.09 eV, 2.92 eV and 3.28 eV between the borophene substrate and Ca, Mg, Na or Li adatom, respectively. Then the adatoms intercalation on the borophene was simulated using a step by step approach. Based on the Bader charge analysis results we acquired remarkably high charge capacities of around 800 mAh/g, 1960 mAh/g, 1380 mAh/g and 1720 mAh/g for Ca, Mg, Na or Li atoms adsorption on borophene, respectively. We predicted that adatoms adsorption on borophene induce considerable deformation and lattice distortion. Nevertheless, using the AIMD simulations at room temperature it was confirmed that the borophene sheets can well retrieve their original structure after the ions separations. In addition, based on the AIMD results we found that borophene films can withstand relatively high temperatures of around 750 K. Moreover, our electronic density of states results also verified that the pristine and charged borophene films present metallic behaviour which is a highly desirable factor for the application of borophene as an anode material for rechargeable ion batteries. We also accomplished the nudged elastic band simulations to obtain diffusion pathways and the corresponding energy barriers. Interestingly, we predicted that diffusion of Mg, Na or Li adatom on borophene is extremely fast and it occurs mainly along the furrow of corrugated borophene. Our investigation proposes the borophene as an outstanding candidate to be utilized as an anode material in rechargeable ion batteries not only because of its high charge capacity but also due to its metallic characteristic, high diffusion rates and also acceptable thermal stability.

## Acknowledgment

BM, OR and TR greatly acknowledge the financial support by European Research Council for COMBAT project (Grant number 615132).

## References

(1)   Novoselov, K. S.; Geim, A. K.; Morozov, S. V; Jiang, D.; Zhang, Y.; Dubonos, S. V; Grigorieva, I. V; Firsov, A. A. *Science* **2004**, *306* (5696), 666–669.

(2)   Geim, A. K.; Novoselov, K. S. *Nat. Mater.* **2007**, *6* (3), 183–191.

(3)   Castro Neto, A. H. .; Peres, N. M. R. .; Novoselov, K. S. .; Geim, A. K. .; Guinea, F. *Rev. Mod. Phys.* **2009**, *81* (1), 109–162.

(4)   Lee, C.; Wei, X.; Kysar, J. W.; Hone, J. *Science (80-. ).* **2008**, *321* (18 July 2008), 385–388.

(5)   Balandin, A. A. *Nat. Mater.* **2011**, *10* (8), 569–581.

(6)   Kubota, Y.; Watanabe, K.; Tsuda, O.; Taniguchi, T. *Science* **2007**, *317* (5840), 932–934.

(7)   Song, L.; Ci, L.; Lu, H.; Sorokin, P. B.; Jin, C.; Ni, J.; Kvashnin, A. G.; Kvashnin, D. G.; Lou, J.; Yakobson, B. I.; Ajayan, P. M. *Nano Lett.* **2010**, *10* (8), 3209–3215.

(8)   Aufray, B.; Kara, A.; Vizzini, Ś.; Oughaddou, H.; Ĺandri, C.; Ealet, B.; Le Lay, G. *Appl. Phys. Lett.* **2010**, *96* (18).

(9)   Vogt, P.; De Padova, P.; Quaresima, C.; Avila, J.; Frantzeskakis, E.; Asensio, M. C.; Resta, A.; Ealet, B.; Le Lay, G. *Phys. Rev. Lett.* **2012**, *108* (15).

(10)  Bianco, E.; Butler, S.; Jiang, S.; Restrepo, O. D.; Windl, W.; Goldberger, J. E. *ACS Nano* **2013**, *7* (5), 4414–4421.

(11)  Zhu, F.; Chen, W.; Xu, Y.; Gao, C.; Guan, D.; Liu, C. *arXiv* **2015**, 1–20.

(12)  Geim, a K.; Grigorieva, I. V. *Nature* **2013**, *499* (7459), 419–425.

(13) Wang, Q. H.; Kalantar-Zadeh, K.; Kis, A.; Coleman, J. N.; Strano, M. S. *Nat. Nanotechnol.* **2012**, *7* (11), 699–712.

(14) Radisavljevic, B.; Radenovic, A.; Brivio, J.; Giacometti, V.; Kis, A. *Nat. Nanotechnol.* **2011**, *6*, 147–150.

(15) Mannix, A. J.; Zhou, X.-F.; Kiraly, B.; Wood, J. D.; Alducin, D.; Myers, B. D.; Liu, X.; Fisher, B. L.; Santiago, U.; Guest, J. R.; Yacaman, M. J.; Ponce, A.; Oganov, A. R.; Hersam, M. C.; Guisinger, N. P. *Science (80-. ).* **2015**, *350* (6267), 1513–1516.

(16) Zhou, X. F.; Dong, X.; Oganov, A. R.; Zhu, Q.; Tian, Y.; Wang, H. T. *Phys. Rev. Lett.* **2014**, *112* (8).

(17) Zhang, Z.; Yang, Y.; Gao, G.; Yakobson, B. I. *Angew. Chemie* **2015**, *127* (44), 13214–13218.

(18) Zhang, Q.; Wang, Y.; Seh, Z. W.; Fu, Z.; Zhang, R.; Cui, Y. *Nano Lett.* **2015**, *15* (6), 3780–3786.

(19) Bonaccorso, F.; Colombo, L.; Yu, G.; Stoller, M.; Tozzini, V.; Ferrari, a. C.; Ruoff, R. S.; Pellegrini, V. *Science (80-. ).* **2015**, *347* (6217), 1246501–1246501.

(20) Bruce, P. G.; Scrosati, B.; Tarascon, J.-M. *Angew. Chem. Int. Ed. Engl.* **2008**, *47*, 2930–2946.

(21) Tritsaris, G. A.; Kaxiras, E.; Meng, S.; Wang, E. *Nano Lett.* **2013**, *13* (5), 2258–2263.

(22) Kasavajjula, U.; Wang, C.; Appleby, A. J. *Journal of Power Sources.* 2007, pp 1003–1039.

(23) Jeong, G.; Kim, Y.-U.; Kim, H.; Kim, Y.-J.; Sohn, H.-J. *Energy Environ. Sci.* **2011**, *4* (6), 1986.

(24) Park, M. H.; Kim, M. G.; Joo, J.; Kim, K.; Kim, J.; Ahn, S.; Cui, Y.; Cho, J. *Nano Lett.* **2009**, *9* (11), 3844–3847.

(25) Chan, T. L.; Chelikowsky, J. R. *Nano Lett.* **2010**, *10* (3), 821–825.

(27) Hwang, H.; Kim, H.; Cho, J. *Nano Lett.* **2011**, *11* (11), 4826–4830.

(28) Withers, F.; Del Pozo-Zamudio, O.; Mishchenko, A.; Rooney, a. P.; Gholinia, A.; Watanabe, K.; Taniguchi, T.; Haigh, S. J.; Geim, a. K.; Tartakovskii, a. I.; Novoselov, K. S. *Nat. Mater.* **2015**, *14* (February), 301–306.

(29) Caldwell, J. D.; Novoselov, K. S. *Nat Mater* **2015**, *14* (4), 364–366.

(31) Hörmann, N. G.; Jäckle, M.; Gossenberger, F.; Roman, T.; Forster-Tonigold, K.; Naderian, M.; Sakong, S.; Groß, A. *J. Power Sources* **2015**, *275*, 531–538.

(32) Legrain, F.; Malyi, O.; Manzhos, S. *J. Power Sources* **2015**, *278*, 197–202.

(33) Zhang, X.; Lu, Z.; Fu, Z.; Tang, Y.; Ma, D.; Yang, Z. *J. Power Sources* **2015**, *276*, 222–229.

(34) Moriwake, H.; Gao, X.; Kuwabara, A.; Fisher, C. A. J.; Kimura, T.; Ikuhara, Y. H.; Kohama, K.; Tojigamori, T.; Ikuhara, Y. *J. Power Sources* **2015**, *276*, 203–207.

(39) Kresse, G.; Furthm??ller, J. *Comput. Mater. Sci.* **1996**, *6* (1), 15–50.

(40) Kresse, G.; Furthmüller, J. *Phys. Rev. B* **1996**, *54* (16), 11169–11186.

(41) Perdew, J.; Burke, K.; Ernzerhof, M. *Phys. Rev. Lett.* **1996**, *77* (18), 3865–3868.

(42) Grimme, S. *J. Comput. Chem.* **2006**, *27* (15), 1787–1799.

(43) Bučko, T.; Hafner, J.; Lebègue, S.; Ángyán, J. G. *J. Phys. Chem. A* **2010**, *114* (43), 11814–11824.

(44) Kresse, G. *Phys. Rev. B* **1999**, *59* (3), 1758–1775.

(45) Tang, W.; Sanville, E.; Henkelman, G. *J. Phys. Condens. Matter* **2009**, *21* (8), 084204.

(46) Henkelman, G.; Uberuaga, B. P.; J??nsson, H. *J. Chem. Phys.* **2000**, *113* (22), 9901–9904.

(47) Chadi, D. J.; Cohen, M. L. *Phys. Rev. B* **1973**, *8* (12), 5747–5753.

(48) Jiang, H. R.; Lu, Z.; Wu, M. C.; Ciucci, F.; Zhao, T. S. *Nano Energy* **2016**, *23*, 97–104.

(49) Momma, K.; Izumi, F. *J. Appl. Crystallogr.* **2011**, *44* (6), 1272–1276.